\documentclass[a4paper]{article}
\pdfoutput=1
\usepackage{graphicx,wrapfig}
\usepackage{lipsum}
\usepackage{amsmath,amssymb,mathrsfs}
\usepackage[hidelinks]{hyperref}
\usepackage{url}
\usepackage{mathtools}
\usepackage{changepage}
\usepackage{multirow}
\usepackage{wrapfig}
\usepackage{subfig}
\usepackage{xcolor}
\usepackage{epsfig,enumerate,amsmath,amsfonts,amssymb,amsthm,mathrsfs,ifpdf}
\usepackage{indentfirst,relsize}
\usepackage{setspace,graphicx}
\usepackage{latexsym}
\usepackage[all]{xy}
\usepackage[margin = 2.50cm]{geometry}
\usepackage[T1]{fontenc}
\usepackage[utf8]{inputenc} 

\usepackage[linesnumbered,ruled,vlined]{algorithm2e}

\usepackage{color}

\usepackage[colorinlistoftodos]{todonotes}

\usepackage{authblk} 

\usepackage{orcidlink}

\usepackage{enumitem}

\usepackage{tikz}
\usetikzlibrary{arrows.meta}

\newcommand{\pcf}{\chi_{\mathrm{pcf}}}

\newcommand{\remove}[1]{}

\newtheorem{theorem}{Theorem}
\newtheorem{lemma}[theorem]{Lemma}

\newtheorem{corollary}[theorem]{Corollary}

\newcounter{Case}[theorem]

\newcounter{Ca}[theorem]

\newcounter{NCa}[theorem]

\usepackage{xcolor}
\allowdisplaybreaks

\title{Proper Conflict-Free Coloring of Cocomparability, Co-Bipartite, and Distance-Hereditary Graphs}

\author[1]{Gautam K. Das\footnote{gkd@iitg.ac.in}}
\author[1]{Kamal Santra \orcidlink{0009-0006-5997-1452} \footnote{kamal.7.2013@gmail.com, kamal.santra@iitg.ac.in}}
\affil[1]{Department of Mathematics\\
	
	Indian Institute of Technology Guwahati\\
	
	Guwahati, 781039, Assam, India}
\date{}
	
\begin{document}

\maketitle
\begin{abstract}
	A proper conflict-free coloring is a proper vertex coloring in which every
	non-isolated vertex has a neighbor whose color occurs exactly once in its
	open neighborhood. We prove that deciding whether $\omega(G)$ colors
	suffice is NP-complete on connected cocomparability graphs of diameter
	two and independence number three. The reduction supplies a partition into
	three cliques, a maximum clique, an optimal ordinary coloring, and a
	dominating edge. Every constructed graph needs either $\omega(G)$ or
	$\omega(G)+1$ PCF colors. The result also gives NP-completeness on
	AT-free graphs. For the co-bipartite subclass, we give a complete matching
	characterization and an $O(n^6)$-time algorithm for an optimal coloring.
	The cocktail-party graphs are exactly the co-bipartite graphs requiring
	two more colors than their ordinary chromatic number; every other
	co-bipartite graph needs at most one additional color. The matching method
	extends to all graphs with independence number at most two, giving an
	$O(n^6\log n)$-time optimization algorithm. Finally, for distance-hereditary
	graphs, we decide whether $k$ colors suffice in $O(n+m+nk^{10})$ time and
	compute an optimal coloring in $O(n^{11}+m)$ time without fixing the
	palette size. The algorithm uses a binary decomposition and groups colors
	according to five possible behaviors at the boundary of a subgraph. 
\end{abstract}



\section{Introduction}
\label{sec:introduction}

A proper conflict-free coloring is a proper vertex coloring in which
every nonisolated vertex has a neighbor whose color appears exactly once
in its open neighborhood. We call such a neighbor a \emph{witness}. On
a three-vertex path, a proper two-coloring gives both ends the same
color and leaves the middle vertex without a witness. A third color
resolves this. When colors represent communication channels, properness
separates adjacent transmitters, and a witness lets a vertex identify
at least one neighbor by its channel.

Fabrici et al studied proper conflict-free
coloring alongside unique-maximum coloring~\cite{FabriciLuzarRindosovaSotak2023}.
Caro et al. investigated its relation to ordinary
coloring~\cite{CaroPetrusevskiSkrekovski2023}. Ahn et al. proved
that PCF $k$-colorability is NP-complete on bipartite graphs for every
fixed $k\ge3$~\cite{AhnImOum2025}. Pradhan and Sharma obtained the same
hardness on perfect elimination bipartite graphs and gave exact
algorithms for several structured classes~\cite{PradhanSharma2026}.

In earlier work, we proved NP-completeness at the clique number on
well-partitioned chordal graphs and gave an exact algorithm for interval
graphs parameterized by the clique number~\cite{SantraStructured2026}.
That work also gave exact algorithms on biconvex and bipartite permutation
graphs and a logarithmic upper bound for convex bipartite graphs.
A subsequent manuscript established hardness at the clique number on
undirected path graphs, even with a supplied representation by paths in
a star, and NP-completeness for each fixed $k\ge4$ on chordal bipartite
graphs~\cite{SantraPathChordalBipartite2026}. It also gave a constructive
$\omega(G)+1$ bound for strongly chordal graphs and exact algorithms for
several of their subclasses, including a linear-time algorithm for
strictly chordal graphs.

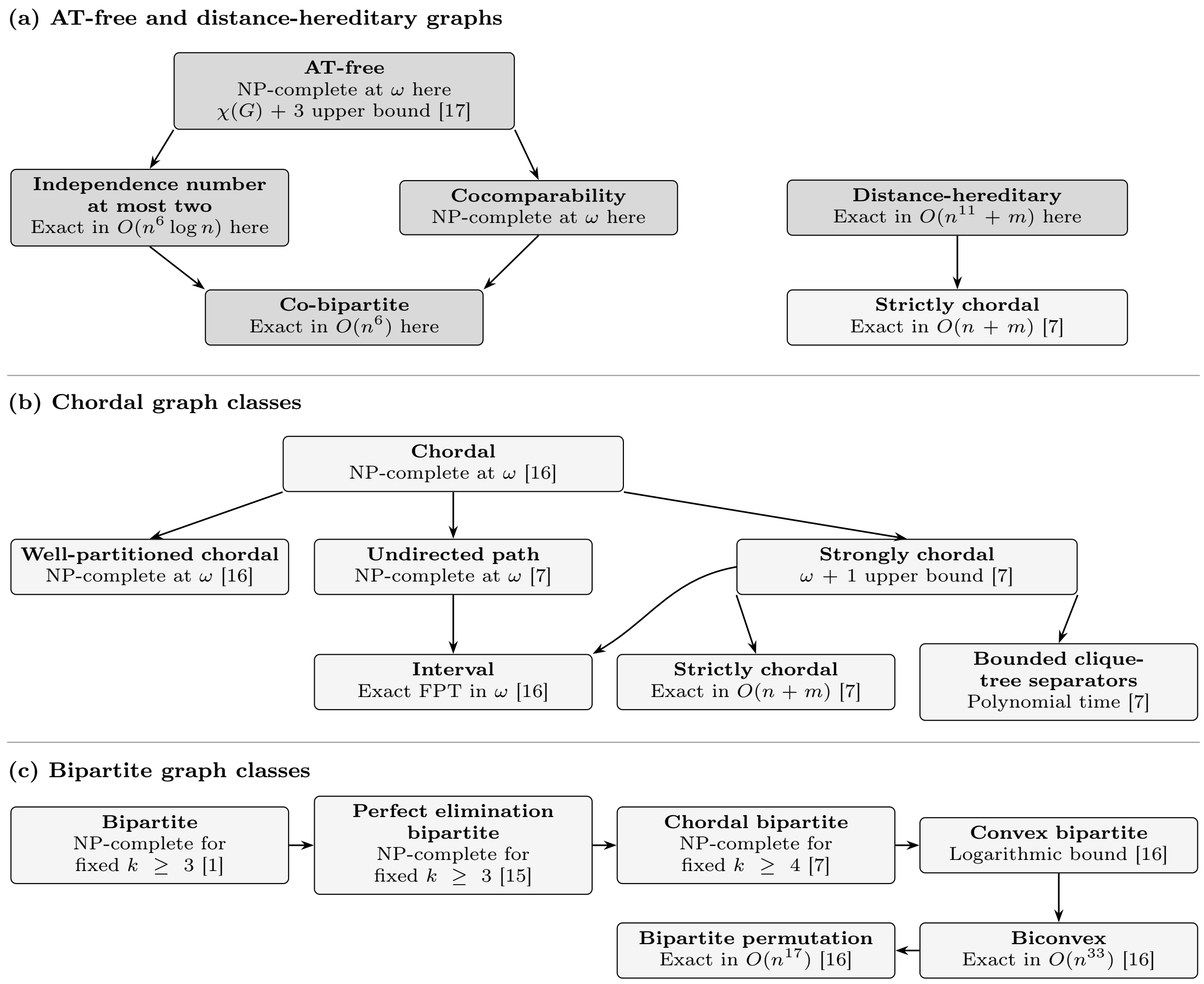
\begin{figure}[h]
\centering
\begin{tikzpicture}[
  x=1cm,y=1cm,
  box/.style={draw,rounded corners=2pt,align=center,font=\scriptsize,
    text width=3.43cm,inner xsep=3pt,inner ysep=3pt,fill=black!4},
  current/.style={box,fill=black!15},
  wide/.style={text width=4.25cm},
  branch/.style={-{Stealth[length=1.7mm,width=1.2mm]},semithick},
  panel/.style={anchor=west,font=\footnotesize\bfseries,inner sep=0pt}
]
  \node[panel] at (0,0) {(a) AT-free and distance-hereditary graphs};
  \node[current,wide] (at) at (4.42,-0.94) {\textbf{AT-free}\\
    NP-complete at $\omega$ here\\
    $\chi(G)+3$ upper bound~\cite{SharmaPaulPandey2026}};
  \node[current] (alpha) at (1.87,-2.47)
    {\textbf{Independence number}\\\textbf{at most two}\\
     Exact in $O(n^6\log n)$ here};
  \node[current] (coco) at (6.97,-2.47) {\textbf{Cocomparability}\\
    NP-complete at $\omega$ here};
  \node[current] (cobi) at (4.42,-3.91) {\textbf{Co-bipartite}\\
    Exact in $O(n^6)$ here};
  \node[current,wide] (dh) at (12.46,-2.47) {\textbf{Distance-hereditary}\\
    Exact in $O(n^{11}+m)$ here};
  \node[box,wide] (strictdh) at (12.46,-3.91) {\textbf{Strictly chordal}\\
    Exact in $O(n+m)$~\cite{SantraPathChordalBipartite2026}};
  \draw[branch] (at.south west) -- (alpha.north);
  \draw[branch] (at.south east) -- (coco.north);
  \draw[branch] (alpha.south) -- (cobi.north west);
  \draw[branch] (coco.south) -- (cobi.north east);
  \draw[branch] (dh.south) -- (strictdh.north);

  \draw[black!40] (0,-4.67) -- (15.64,-4.67);
  \node[panel] at (0,-5.04) {(b) Chordal graph classes};
  \node[box,wide] (chordal) at (5.85,-5.83) {\textbf{Chordal}\\
    NP-complete at $\omega$~\cite{SantraStructured2026}};
  \node[box] (wpc) at (1.87,-7.18) {\textbf{Well-partitioned chordal}\\
    NP-complete at $\omega$~\cite{SantraStructured2026}};
  \node[box] (path) at (5.85,-7.18) {\textbf{Undirected path}\\
    NP-complete at $\omega$~\cite{SantraPathChordalBipartite2026}};
  \node[box,wide] (strong) at (11.80,-7.18) {\textbf{Strongly chordal}\\
    $\omega+1$ upper bound~\cite{SantraPathChordalBipartite2026}};
  \node[box] (interval) at (5.85,-8.69) {\textbf{Interval}\\
    Exact FPT in $\omega$~\cite{SantraStructured2026}};
  \node[box] (strict) at (9.82,-8.69) {\textbf{Strictly chordal}\\
    Exact in $O(n+m)$~\cite{SantraPathChordalBipartite2026}};
  \node[box] (sep) at (13.79,-8.69) {\textbf{Bounded clique-tree separators}\\
    Polynomial time~\cite{SantraPathChordalBipartite2026}};
  \draw[branch] (chordal.south west) -- (wpc.north);
  \draw[branch] (chordal.south) -- (path.north);
  \draw[branch] (chordal.south east) -- (strong.north);
  \draw[branch] (path.south) -- (interval.north);
  \draw[branch] (strong.west) to[out=190,in=30] (interval.north east);
  \draw[branch] (strong.south west) -- (strict.north);
  \draw[branch] (strong.south east) -- (sep.north);

  \draw[black!40] (0,-9.48) -- (15.64,-9.48);
  \node[panel] at (0,-9.87) {(c) Bipartite graph classes};
  \node[box] (bip) at (1.87,-10.83) {\textbf{Bipartite}\\
    NP-complete for fixed $k\ge3$~\cite{AhnImOum2025}};
  \node[box] (peb) at (5.85,-10.83) {\textbf{Perfect elimination}\\\textbf{bipartite}\\
    NP-complete for fixed $k\ge3$~\cite{PradhanSharma2026}};
  \node[box] (cb) at (9.82,-10.83) {\textbf{Chordal bipartite}\\
    NP-complete for fixed $k\ge4$~\cite{SantraPathChordalBipartite2026}};
  \node[box] (conv) at (13.79,-10.83) {\textbf{Convex bipartite}\\
    Logarithmic bound~\cite{SantraStructured2026}};
  \node[box] (biconv) at (13.79,-12.21) {\textbf{Biconvex}\\
    Exact in $O(n^{33})$~\cite{SantraStructured2026}};
  \node[box] (bp) at (9.82,-12.21) {\textbf{Bipartite permutation}\\
    Exact in $O(n^{17})$~\cite{SantraStructured2026}};
  \draw[branch] (bip.east) -- (peb.west);
  \draw[branch] (peb.east) -- (cb.west);
  \draw[branch] (cb.east) -- (conv.west);
  \draw[branch] (conv.south) -- (biconv.north);
  \draw[branch] (biconv.west) -- (bp.east);
\end{tikzpicture}
\caption{Selected graph-class inclusions and results for proper conflict-free
coloring. Arrows point from a class to a subclass. Darker boxes mark results
of this paper. The symbol $\omega$ denotes the clique number, and
NP-completeness at $\omega$ refers to the decision whether $\omega(G)$ colors
suffice. Strictly chordal graphs appear in panels (a) and (b) to show both
superclasses. In panel (b), the separator-size bound is fixed.}
\label{fig:class-hierarchy}
\end{figure}

Figure~\ref{fig:class-hierarchy} shows how the classes studied here
and in the two companion manuscripts are related.

Here we first consider cocomparability graphs, whose complements admit
transitive orientations. These graphs are AT-free
\cite{CorneilOlariuStewart1997}. Sharma et al. showed that
$\pcf(G)\le\chi(G)+3$ for AT-free graphs and gave exact results for
several structured classes~\cite{SharmaPaulPandey2026}. The additive
bound leaves few possible values for the optimum. We show that it is
already hard to decide between $\omega(G)$ and $\omega(G)+1$, even
when the reduction supplies an optimal ordinary coloring and a
dominating edge.

For co-bipartite graphs, a partition into two cliques lets us represent
each proper coloring by a matching of cross nonedges. The choice of
matching determines which vertices have witnesses. Jiménez et
al.~\cite{JimenezEtAl2026} established the cocktail-party lower bound.
We characterize when that bound is attained within the co-bipartite
class and use the matching representation to find an optimal coloring.
The same representation applies whenever the independence number is
at most two, because then each color class is a singleton or a
nonadjacent pair. A related matching approach for strong odd coloring
appears in work of Caro et al.~\cite{CaroEtAlOdd2026}. Strong odd
coloring requires every color present in a neighborhood to have odd
multiplicity.

Our last setting is distance-hereditary graphs. Their construction
by pendant additions and true or false twins gives a binary
decomposition in which the exposed vertices of each subgraph have
the same outside neighborhood~\cite{BandeltMulder1986}. This lets us
group colors with the same effect on the exposed vertices and
optimize over an unrestricted palette. The lower bound for
open-neighborhood conflict-free coloring of bipartite
distance-hereditary graphs also applies to PCF coloring, since a
proper conflict-free coloring is a feasible conflict-free coloring
after dropping the properness requirement~\cite{BhyravarapuEtAl2024}.

Strictly chordal graphs form a proper subclass of distance-hereditary
graphs. A strictly chordal graph is obtained from a block graph by
replacing each vertex with a nonempty clique and joining the
replacement cliques of adjacent vertices completely
\cite{DumasEtAl2021}. To see the inclusion, root each block-cut tree
at a block. Build the root clique by adding true twins to one vertex.
For each child block, add one pendant neighbor to its parent cut
vertex and then add the other new vertices of the block as true twins
of that pendant vertex. The clique replacements are further true-twin
additions. The distance-hereditary construction theorem now gives
the inclusion~\cite{BandeltMulder1986}. The four-cycle belongs to the
larger class, as it can be built from an edge by adding a false twin
of each endpoint, and strictly chordal graphs are chordal. The earlier
linear-time algorithm exploits the weighted blocks of the true-twin
quotient~\cite{SantraPathChordalBipartite2026}. The algorithm here
handles the full distance-hereditary class.

\medskip
\noindent\textbf{Our results}\par
Write $n=|V(G)|$ and $m=|E(G)|$.
\begin{enumerate}[leftmargin=*,itemsep=0.45em]
  \item \textbf{Cocomparability and AT-free graphs.}
  Deciding whether $\pcf(G)=\omega(G)$ is NP-complete on connected
  cocomparability graphs of diameter two and independence number
  three. The reduction supplies a partition into three cliques,
  a maximum clique, an optimal ordinary coloring, and a dominating
  edge. Each constructed graph has optimum $\omega(G)$ or
  $\omega(G)+1$. This also gives NP-completeness on AT-free graphs
  (Theorem~\ref{thm:cocomparability-hardness} and
  Corollary~\ref{cor:atfree-hardness}).
  \item \textbf{Co-bipartite graphs and independence number at most two.}
  We give a matching characterization and an $O(n^6)$-time algorithm
  for optimal coloring on co-bipartite graphs. Among these graphs,
  exactly the cocktail-party graphs require $\chi(G)+2$ colors.
  Every other co-bipartite graph needs at most $\chi(G)+1$. On all
  graphs with independence number at most two, our matching method
  gives an $O(n^6)$-time decision algorithm and an $O(n^6\log n)$-time
  optimization algorithm (Section~\ref{sec:cobipartite}).
  \item \textbf{Distance-hereditary graphs.}
  We decide PCF $k$-colorability in $O(n+m+nk^{10})$ time and find an
  optimal coloring in $O(n^{11}+m)$ time. A histogram of five color
  types represents how the colors used in a subgraph interact with
  its outside neighbors (Theorem~\ref{thm:dh-exact}).
\end{enumerate}

Section~\ref{sec:preliminaries} introduces the notation and graph
classes. Section~\ref{sec:atfree-hardness} proves hardness on
cocomparability graphs and derives the AT-free corollary.
Section~\ref{sec:cobipartite} develops the matching algorithms,
and Section~\ref{sec:dh} gives the distance-hereditary algorithm.
Section~\ref{sec:conclusion} discusses open questions.

\section{Preliminaries}
\label{sec:preliminaries}

All graphs are finite, simple, and undirected. Write $n=|V(G)|$ and
$m=|E(G)|$. The open neighborhood of a
vertex $v$ is denoted by $N_G(v)$, and
$N_G[v]=N_G(v)\cup\{v\}$ is its closed neighborhood. We write $G[X]$ for
the subgraph induced by $X$. A \emph{proper conflict-free coloring}, or
\emph{PCF coloring}, is a proper coloring $c$ such that every
non-isolated vertex $v$ has a color occurring exactly once on $N_G(v)$.
The minimum number of colors is denoted by $\pcf(G)$. Isolated vertices
need a color but do not need a witness. We set the value of the empty
graph to zero.

For every nonempty graph, $\chi(G)\leq\pcf(G)\leq n$. For a disconnected
graph, the value is the maximum over its components because the same
palette may be reused. A color class is an independent set. We write
$\omega(G)$ and $\alpha(G)$ for the clique number and independence number.

An orientation is \emph{transitive} if, whenever $x\to y$ and
$y\to z$ are arcs, the edge $xz$ is present and oriented $x\to z$.
A graph is a \emph{comparability graph} if it has a transitive orientation.
A graph is a \emph{cocomparability graph} if its complement is a
comparability graph.

An \emph{asteroidal triple} is an independent set of three vertices such
that every two are joined by a path avoiding the closed neighborhood of
the third. A graph is \emph{AT-free} if it has no asteroidal triple.
Cocomparability graphs are AT-free \cite{CorneilOlariuStewart1997}.
A graph with independence number at most two is also AT-free, since it
has no independent triple.

Only three multiplicities are relevant in a neighborhood: zero, one, and
at least two. We therefore use the truncation
$\tau(x)=\min\{2,x\}$. A truncated count of one supplies a witness. Unless
stated otherwise, the input palette size is part of the instance. 

We use \textsc{Exact Cover by 3-Sets} (X3C) in the hardness proof. An
instance consists of a universe $U$ with $|U|=3q$ and a family of
three-element subsets of $U$; the question is whether exactly $q$ sets
partition $U$. X3C is NP-complete \cite{GareyJohnson1979}.

\section{NP-completeness on cocomparability graphs}
\label{sec:atfree-hardness}

Sharma et al.~\cite{SharmaPaulPandey2026} proved that every
AT-free graph satisfies $\chi_{\mathrm{pcf}}(G)\leq\chi(G)+3$.
This bound does not decide whether a given palette suffices. We prove
NP-completeness on cocomparability graphs and then obtain AT-free
hardness as a corollary. The number of colors is part of the input.
Our construction has a partition into three cliques and an explicit
optimal ordinary coloring, so the difficulty comes from the witness
requirement.

\begin{theorem}
\label{thm:cocomparability-hardness}
PCF Colorability is NP-complete on connected cocomparability graphs of
diameter two and independence number three. The result holds when a
partition into three cliques is supplied, and the target number of colors
is $k=\chi(G)=\omega(G)$. For the graphs in the reduction, $\chi_{\mathrm{pcf}}(G)$ is either
$\omega(G)$ or $\omega(G)+1$. Distinguishing these two values is
NP-complete.
\end{theorem}

\begin{proof}
A proposed coloring can be checked in polynomial time, so PCF Colorability
belongs to NP. For the equality question, a PCF coloring and a clique with
the same number of vertices certify
$\chi_{\mathrm{pcf}}(G)=\omega(G)$.

We reduce from Exact Cover by 3-Sets (X3C)~\cite{GareyJohnson1979}.
Let $U=\{u_1,\ldots,u_{3q}\}$ and let
$\mathcal S=\{S_1,\ldots,S_m\}$ consist of three-element subsets of $U$.
We ask whether $q$ members of $\mathcal S$ partition $U$.
We may restrict to $q\geq1$ and $m\geq q$. Instances with $q=0$ are
yes-instances, while those with $m<q$ and $q\geq1$ are no-instances.
They can be replaced by fixed X3C yes- and no-instances satisfying the
restriction. For example, a yes-instance has universe $\{1,2,3\}$ and
its single three-element set; a no-instance has universe
$\{1,\ldots,6\}$ and the two sets $\{1,2,3\},\{1,2,4\}$.

For each set $S_i$, introduce two ports $a_i,b_i$. Add two anchors $s,t$
and make
$B=U\mathbin{\dot\cup}\{a_i,b_i:i\in[m]\}
  \mathbin{\dot\cup}\{s,t\}$
a clique. Put $K=|B|=3q+2m+2$.

Take a second clique $A$ of order $K$. Its vertices are
$\widehat u$ for $u\in U$, the pairs $\widehat a_i,\widehat b_i$, and
the two anchors $\widehat s,\widehat t$. The hats distinguish vertices
of $A$ from their namesakes in $B$. We identify each element
$u\in U$ with its corresponding vertex in $B$. Consequently, for
each $i\in[m]$, the set $S_i$ appearing in the table denotes
the three element vertices of $B$ corresponding to the members
of $S_i$, rather than an individual vertex.

For a vertex $x\in A$, let $\Lambda(x)$ specify its
nonneighbors in $B$, as follows:
\begin{center}
\begin{tabular}{l|l}
\hline
vertex of $A$ & nonneighbors in $B$\\
\hline
$\widehat s,\widehat t$ & $\{s,t\}$\\
$\widehat u$, for $u\in U$ & $\{u,s\}$\\
$\widehat a_i,\widehat b_i$, for $i\in[m]$
  & $\{a_i,b_i,s\}\cup S_i$\\
\hline
\end{tabular}
\end{center}
Every other pair with one endpoint in $A$ and one in $B$ is an edge.
The sets $\Lambda(x)$ only define graph nonedges; there are no lists or
preassigned colors in the instance.

Finally, add a clique $Z$ of order $2m-q$. Make it complete to
$U\cup\{s,t\}$, anticomplete to the ports in $B$, and anticomplete to
$A$. These are all the edges of the graph, denoted by $G$.
Figure~\ref{fig:atfree-x3c} shows the construction. All three cliques
are nonempty, and $|V(G)|=2K+2m-q=5q+6m+4$, so the reduction has
polynomial size.

\begin{figure}[t]
\centering
\begin{tikzpicture}[
  bag/.style={draw,thick,rounded corners,align=center,fill=white,
              minimum height=1.45cm},
  every node/.style={font=\small},
  baglink/.style={thick}
]
\node[bag,minimum width=3.6cm] (A) at (-5,0)
  {$A$: a $K$-clique\\
   $\widehat U,\ \widehat a_i,\widehat b_i,\ \widehat s,\widehat t$};
\node[bag,minimum width=3.9cm] (B) at (0,0)
  {$B$: a $K$-clique\\
   $U,\ a_i,b_i,\ s,t$};
\node[bag,minimum width=2.7cm] (Z) at (4.7,0)
  {$Z$: a clique\\$|Z|=2m-q$};
\draw[baglink] (A.east) -- (B.west);
\node[align=center,font=\scriptsize] at (-2.55,1.35)
  {all edges except\\$xv$ with $v\in\Lambda(x)$};
\draw[baglink] (B.east) -- (Z.west);
\node[align=center,font=\scriptsize] at (2.5,1.35)
  {complete from $U\cup\{s,t\}$\\
   no edges from the ports};
\node[align=center,font=\small] at (-0.2,-1.45)
  {$A$ and $Z$ are anticomplete. The sets $\Lambda(x)$ are given in the table.};
\end{tikzpicture}
\caption{The three-clique construction. All vertices inside a displayed
bag are pairwise adjacent. The omitted pairs between $A$ and $B$ force
the colors on $A$ while allowing the element vertices to test the
missing port colors on $Z$.}
\label{fig:atfree-x3c}
\end{figure}

We first verify the graph class. Orient the complement from $A$ to $B$
on every pair specified by $\Lambda$, from $A$ to $Z$ on every pair,
and from each port of $B$ to every vertex of $Z$. These are all the
edges of the complement. The only directed paths of length two have
the form $A\to B_{\rm port}\to Z$, and their endpoints are joined by
an arc from $A$ to $Z$. Thus the orientation is transitive, and $G$
is a cocomparability graph.

There is also a direct check that $G$ is AT-free. An independent triple
must contain one vertex from each of $A,B,Z$. Let $b\in B$ be its
$B$-vertex. Since $B$ is a clique, $G-N[b]$ contains no vertex of $B$.
Moreover, $A\setminus N[b]$ and $Z\setminus N[b]$ are anticomplete
because $A$ and $Z$ are anticomplete. Hence, the $A$-vertex and the
$Z$-vertex of the triple cannot be joined by a path avoiding $N[b]$.
The triple is therefore not asteroidal.

The partition shows $\alpha(G)\leq3$. For any $i$ and any $z\in Z$,
the vertices $\widehat a_i,a_i,z$ are independent, so $\alpha(G)=3$.
Every nonanchor vertex of $A$ is adjacent to $t\in B$, while the two
anchors of $A$ are adjacent to every element vertex in $U$. All these
vertices of $B$ are adjacent to every vertex of $Z$. Thus, every
$A$--$Z$ pair has distance two. An $A$--$B$ nonedge has a length-two
path through a neighbor of its $A$ endpoint in the clique $B$.
A $B$--$Z$ nonedge has a length-two path through $s$. Pairs in the same
bag are adjacent. Therefore $G$ is connected and has diameter two.

A proper $K$-coloring always exists. Give the vertices of $B$ distinct
colors, and name each color by its vertex in $B$. Give every vertex of
$A$ its namesake color. Color $Z$ with any $2m-q$ distinct port colors.
Each chosen color is allowed by the corresponding nonedge, and
$A,Z$ are anticomplete. Since $B$ is a $K$-clique, we obtain
$\chi(G)=\omega(G)=K$.

Now consider any proper $K$-coloring. Rename the colors using $B$.
The two anchors in $A$ can only use $s,t$, so they use both colors.
Each $\widehat u$ can only use $u$ or $s$; since color $s$ is already
used in the clique $A$, it must use $u$. The vertices
$\widehat a_i,\widehat b_i$ can only use colors in
$\{a_i,b_i,s\}\cup S_i$. The element colors and $s$ are already used
in $A$, so this pair uses $a_i,b_i$, in either order.
Consequently, the colors on $A$ are fixed up to the anchor swap and
the swaps within port pairs.

The clique $Z$ can only use port colors. It uses $2m-q$ of them, each
once. Let $T$ be the set of $q$ port colors missing from $Z$.

Fix an element vertex $u\in U\subseteq B$. The following table gives the
multiplicity of each color in its neighborhood. A port $p\in\{a_i,b_i\}$
is called incident with $u$ when $u\in S_i$.
\begin{center}
\begin{tabular}{l|c}
\hline
color & multiplicity in $N(u)$\\
\hline
$u$ & $0$\\
$v\in U\setminus\{u\}$ & $2$\\
$s$ or $t$ & $2$\\
a port $p$ not incident with $u$ & $2+\mathbf{1}_{p\notin T}$\\
a port $p$ incident with $u$ & $1+\mathbf{1}_{p\notin T}$\\
\hline
\end{tabular}
\end{center}
Here $\mathbf{1}_{p\notin T}$ is one when $p\notin T$ and zero otherwise.
Every color different from $u$ appears once in $B\setminus\{u\}$.
For an element color $v\neq u$, its copy $\widehat v$ is adjacent to
$u$, and both anchors of $A$ are adjacent to $u$. These give the second
occurrence of every element and anchor color. A port color has its
copy in the corresponding pair of $A$. This copy is nonadjacent to $u$
exactly when the set contains $u$, independently of the possible swap
within the pair. Finally, $Z$ contributes a port color exactly when it
does not belong to $T$. The color $u$ is absent from the neighborhood
by properness.

It follows that $u$ has a uniquely colored neighbor if and only if
$T$ contains a port belonging to a set that contains $u$.
No other vertex imposes a further condition:
\begin{itemize}
\item Every nonanchor vertex of $A$ is witnessed by the vertex of $A$
colored $s$. This color occurs once in $A$, its vertex in $B$ is a
nonneighbor, and there are no edges from $A$ to $Z$.
The two anchors of $A$ witness one another.
\item A port $a_i$ in $B$ is witnessed by color $b_i$, and conversely.
Its occurrence in $A$ lies in the corresponding pair and is a
nonneighbor, while all of $Z$ is anticomplete to the ports.
\item The anchors $s,t$ in $B$ witness one another. The occurrence of
the other anchor color in $A$ is a nonneighbor, and $Z$ uses no anchor
color.
\item Every vertex of $Z$ is witnessed by $s\in B$. No other neighbor
has color $s$.
\end{itemize}
Thus the proper coloring is PCF exactly when the omitted port colors
cover all elements through their sets.

If $\{S_i:i\in I\}$ is an exact cover, take
$T=\{a_i:i\in I\}$. Use the canonical coloring of $A,B$ above and color
$Z$ with all port colors outside $T$. Every element has an incident
omitted port, and the preceding witness checks handle every other
vertex. This gives a PCF $K$-coloring.

Conversely, suppose that a PCF $K$-coloring exists. The sets whose ports
occur in $T$ cover $U$. There are at most $q$ such sets, each of order
three. Since $|U|=3q$, they must be exactly $q$ sets with no overlap.
In particular, $T$ cannot spend two of its $q$ colors on the same set.
These sets form an exact cover. This proves NP-hardness with target
$k=K=\chi(G)=\omega(G)$.

It remains to show that one extra color always suffices. Start with the
canonical proper $K$-coloring and any permitted coloring of $Z$. Recolor
$\widehat s$ with a fresh color. It now witnesses every other vertex
of $A$ and every element or port vertex of $B$. The vertex $\widehat s$
is witnessed by $\widehat t$, since $t\in B$ is its nonneighbor.
In $B$, the anchor $s$ is still witnessed by $t$, while $t$ is
witnessed by $s$, whose color now occurs only at its vertex in $B$.
Every vertex of $Z$ is also witnessed by $s$.
The coloring is proper and uses $K+1$ colors.

Hence, every output graph has PCF chromatic number $K$ or $K+1$, and
it has the lower value exactly for X3C yes-instances. This completes
the proof.
\end{proof}

\begin{corollary}
\label{cor:atfree-hardness}
PCF Colorability is NP-complete on connected AT-free graphs of diameter
two and independence number three, even when a partition into three
cliques is supplied and the target is $k=\chi(G)=\omega(G)$.
The hardness holds for graphs with
$\chi_{\mathrm{pcf}}(G)\in\{\omega(G),\omega(G)+1\}$.
\end{corollary}

\begin{proof}
Every cocomparability graph is AT-free, so
Theorem~\ref{thm:cocomparability-hardness} gives the result.
\end{proof}

\begin{corollary}
\label{cor:atfree-supplied-information}
The hardness in Theorem~\ref{thm:cocomparability-hardness} remains when the input
also includes a maximum clique, an optimal ordinary coloring, and a
dominating edge.
\end{corollary}

\begin{proof}
Supply the clique $B$, the canonical proper $K$-coloring, and the edge
$tu$ for any $u\in U$. The vertex $t$ is adjacent to every nonanchor
vertex of $A$ and every vertex of $Z$. The vertex $u$ is adjacent to
both anchors of $A$, while the clique $B$ is covered by the endpoints
of the edge. Thus, $tu$ is a dominating edge. The earlier proof gives
$|B|=\chi(G)=\omega(G)=K$, so the other two supplied objects are optimal.
\end{proof}

The palette size is not fixed in this reduction. On a graph of
independence number at most three, a proper $k$-coloring can cover at most
$3k$ vertices. For each fixed $k$, we can therefore reject larger graphs
and exhaustively test the remaining ones. The theorem establishes
hardness with $k$ as part of the input.

\section{Co-bipartite graphs and the matching method}
\label{sec:cobipartite}

Theorem~\ref{thm:cocomparability-hardness} establishes hardness on
cocomparability graphs. We now consider co-bipartite graphs, a subclass
on which the exact problem can be solved in polynomial time.
A graph is co-bipartite if its vertices can be partitioned into two cliques
$A,B$. Its complement is bipartite. Orient every edge of the complement
from $A$ to $B$. There is no directed path of length two, so this
orientation is transitive. Thus, every co-bipartite graph is a
cocomparability graph. It also has independence number at most two and
is therefore AT-free.

We give an exact matching characterization of the PCF chromatic number.
The answer is either $\chi(G)$ or $\chi(G)+1$, except for the
cocktail-party graphs $\mathrm{CP}_r=K_{2r}-rK_2$, with $r\geq2$,
which require two additional colors. Their lower bound is known
\cite{JimenezEtAl2026}. The characterization below identifies all the
other cases and gives a polynomial-time exact algorithm.

Let $V(G)=A\mathbin{\dot\cup}B$, where $A,B$ are cliques, and write
$a=|A|$, $b=|B|$.  We may assume $a\geq b$.  The \emph{cross-nonedge
graph} $H$ is the bipartite graph with parts $A$ and $B$ in which $xy$ is an
edge precisely when $x\in A$, $y\in B$, and $xy\notin E(G)$.
A partition into two cliques can be found by testing whether the complement
of $G$ is bipartite.  The cases of an empty graph or an empty part are
immediate, so we assume $a\geq b\geq1$ unless stated otherwise.

\subsection{Colorings represented by matchings}

Every color class of a proper coloring contains at most one vertex from
each clique.  A class of size two therefore corresponds to an edge of $H$,
and these edges form a matching.  Conversely, a matching $M$ in $H$
defines a proper coloring: give the endpoints of each matching edge one
color, and give each unmatched vertex its own new color.  We call this the
\emph{matching coloring} of $M$.  Its unmatched vertices have colors that
occur only once in the whole graph.  It follows directly that
$\chi(G)=a+b-\nu(H)$, where $\nu(H)$ is the maximum matching size.

\begin{lemma}
\label{lem:cobip-unsaturated}
If $\nu(H)<\min\{a,b\}$, then every maximum matching coloring is PCF.
In particular, $\chi_{\mathrm{pcf}}(G)=\chi(G)$.
\end{lemma}

\begin{proof}
Let $M$ be a maximum matching, and let $U_A,U_B$ be its unmatched vertices
in the two parts.  Both sets are nonempty.  No edge of $H$ joins
$U_A$ to $U_B$, since such an edge could be added to $M$. Thus, every
vertex of $U_A$ is adjacent in $G$ to every vertex of $U_B$.

A matched vertex of $A$ sees the private color of any vertex in $U_A$.
An unmatched vertex of $A$ sees another private color in $A$ if
$|U_A|\geq2$.  If it is the only unmatched vertex of $A$, it instead
sees the private color of any vertex in $U_B$.  The same argument applies
to $B$.  Every vertex therefore has a unique neighbor color, and the
matching coloring is PCF.  It uses exactly $\chi(G)$ colors.
\end{proof}

The remaining cases have a maximum matching saturating one part.  The next
observation describes precisely when a vertex fails the PCF condition.

\begin{lemma}
\label{lem:cobip-defects}
Suppose $a>b$, put $q=a-b$, and let $M$ saturate $B$ in $H$.
Let $U\subseteq A$ be the unmatched set of $M$, so $|U|=q$.
In the matching coloring of $M$, the following statements hold.
\begin{enumerate}[label=(\roman*),leftmargin=*]
\item If $q\geq2$, every vertex of $A$ has a witness.  If $q=1$, all
vertices of $A$ except the unique $u\in U$ have a witness, and $u$ has
a witness exactly when $d_H(u)>0$.
\item A nonisolated vertex $y\in B$ fails the PCF condition exactly when
$d_H(y)=q+1$ and $U\subseteq N_H(y)$.
\end{enumerate}
Consequently, whether this coloring is PCF depends on $U$, rather than on
the particular matching saturating $B$ with unmatched set $U$.
\end{lemma}

\begin{proof}
Each unmatched vertex has a private color.  If $q\geq2$, every vertex of
$A$ is adjacent to an unmatched vertex different from itself.  If $q=1$,
this argument covers all of $A$ except its unique unmatched vertex $u$.
For $u$, consider a matched pair $xy$ with $x\in A$, $y\in B$.
The vertex $u$ is adjacent to $x$, and the pair's color occurs once in
$N_G(u)$ precisely when $uy\in E(H)$.  All vertices of $B$ are matched,
so this proves (i).

For (ii), let $xy\in M$.  The color of $x,y$ is absent from $N_G(y)$.
If $y$ is adjacent in $G$ to some $u\in U$, its private color witnesses
$y$.  Suppose there is no such vertex, so $U\subseteq N_H(y)$.
For a different matched pair $x'y'$, the vertex $y$ is adjacent to $y'$
and sees the pair's color exactly once if and only if $yx'\in E(H)$.
Thus $y$ fails precisely when it has no $H$-neighbor among the matched
vertices of $A$ other than its partner $x$.  Together with
$U\subseteq N_H(y)$, this says $N_H(y)=U\cup\{x\}$.
It is equivalent to $d_H(y)=q+1$ and $U\subseteq N_H(y)$, because the
matching partner must lie in $N_H(y)\setminus U$.  Isolated vertices of
$G$ have no PCF requirement and are excluded in (ii).
\end{proof}

\subsection{A complete classification for equal part sizes}

\begin{theorem}
\label{thm:cobip-balanced}
Suppose $|A|=|B|=r\geq2$ and put $\nu=\nu(H)$.  Then:
\begin{enumerate}[label=(\roman*),leftmargin=*]
\item if $\nu<r$, then $\chi_{\mathrm{pcf}}(G)=2r-\nu=\chi(G)$;
\item if $\nu=r$ and $\delta(H)\geq2$, then
$\chi_{\mathrm{pcf}}(G)=r$;
\item if $\nu=r$, $\delta(H)=1$, and $H\neq rK_2$, then
$\chi_{\mathrm{pcf}}(G)=r+1$;
\item if $H=rK_2$, then $G=\mathrm{CP}_r$ and
$\chi_{\mathrm{pcf}}(G)=r+2$.
\end{enumerate}
An optimal coloring can be constructed in polynomial time.
\end{theorem}

\begin{proof}
Part (i) is Lemma~\ref{lem:cobip-unsaturated}.  Now suppose $H$ has a
perfect matching $M$, so $\chi(G)=r$.  Every proper $r$-coloring comes
from a perfect matching of $H$.  For a vertex $x\in A$, its own matching
color is absent from its neighborhood.  Every other color is represented
by a pair $x'y'$, where $x'$ is adjacent to $x$ in the clique $A$.
This color occurs once in $N_G(x)$ exactly when $xy'$ is an edge of $H$.
Thus $x$ has a witness precisely when it has an $H$-neighbor other than
its matching partner.  The same statement holds on $B$.
All vertices are nonisolated in $G$, because $r\geq2$.
It follows that a proper $r$-coloring is PCF exactly when
$\delta(H)\geq2$.  This proves (ii) and the lower bound in (iii).

If $H\neq rK_2$, it has an edge $xy$ outside the chosen perfect matching.
Write $xy_0,x_0y\in M$.  The four vertices are distinct.  Replace these
two matching edges by $xy$, obtaining a matching $M'$ of size $r-1$.
Its unmatched vertices are $x_0$ and $y_0$.  In the matching coloring,
every other vertex of $A$ sees the private color of $x_0$, and every
other vertex of $B$ sees the private color of $y_0$.
The new paired color on $x,y$ witnesses both unmatched vertices:
$x_0$ is adjacent in $G$ to $x$ and nonadjacent to $y$, while $y_0$
is adjacent to $y$ and nonadjacent to $x$.  This gives a PCF coloring
with $r+1$ colors and proves (iii).

Finally, let $H=rK_2$.  A coloring with at most $r+1$ colors would
correspond to a matching of size at least $r-1$.  A perfect matching gives
no witness, as just proved.  A matching of size $r-1$ leaves the endpoints
$x_0,y_0$ of one edge of $H$ unmatched.  They are nonadjacent in $G$,
and every color in either of their neighborhoods occurs twice, once in
each of the other matched pairs.  This coloring is not PCF either.
For the upper bound, retain $r-2$ matching edges and leave two vertices
unmatched in each part.  Every vertex now sees a private color in its own
part.  The resulting coloring uses $(r-2)+4=r+2$ colors.
The matching and the replacements described above can be found in
polynomial time.
\end{proof}

\begin{figure}[t]
\centering
\begin{tikzpicture}[x=0.8cm,y=0.8cm,
  vertex/.style={circle,draw=black,fill=white,inner sep=1.3pt,font=\small}]
\begin{scope}
\foreach \i in {1,2,3}{
  \node[vertex] (pa\i) at (0,1-\i) {$a_\i$};
  \node[vertex] (pb\i) at (2,1-\i) {$b_\i$};
}
\foreach \i/\j in {1/1,1/2,2/2,2/3,3/3,3/1}
  \draw[black] (pa\i)--(pb\j);
\node[align=center,font=\small] at (1,-3.05)
  {$\delta(H)=2$\\$\chi_{\mathrm{pcf}}(G)=3$};
\end{scope}
\begin{scope}[xshift=4.8cm]
\foreach \i in {1,2,3}{
  \node[vertex] (qa\i) at (0,1-\i) {$a_\i$};
  \node[vertex] (qb\i) at (2,1-\i) {$b_\i$};
}
\foreach \i/\j in {1/1,2/2,3/3,1/2}
  \draw[black] (qa\i)--(qb\j);
\node[align=center,font=\small] at (1,-3.05)
  {$\delta(H)=1$, $H\neq3K_2$\\$\chi_{\mathrm{pcf}}(G)=4$};
\end{scope}
\begin{scope}[xshift=9.6cm]
\foreach \i in {1,2,3}{
  \node[vertex] (ra\i) at (0,1-\i) {$a_\i$};
  \node[vertex] (rb\i) at (2,1-\i) {$b_\i$};
}
\foreach \i in {1,2,3}\draw[black] (ra\i)--(rb\i);
\node[align=center,font=\small] at (1,-3.05)
  {$H=3K_2$\\$\chi_{\mathrm{pcf}}(G)=5$};
\end{scope}
\end{tikzpicture}
\caption{The three perfect-matching cases with $|A|=|B|=3$.
The drawn edges belong to $H$: they are the missing cross edges of $G$.
Both parts induce cliques in $G$.  In each panel $\chi(G)=3$.}
\label{fig:cobip-three-values}
\end{figure}
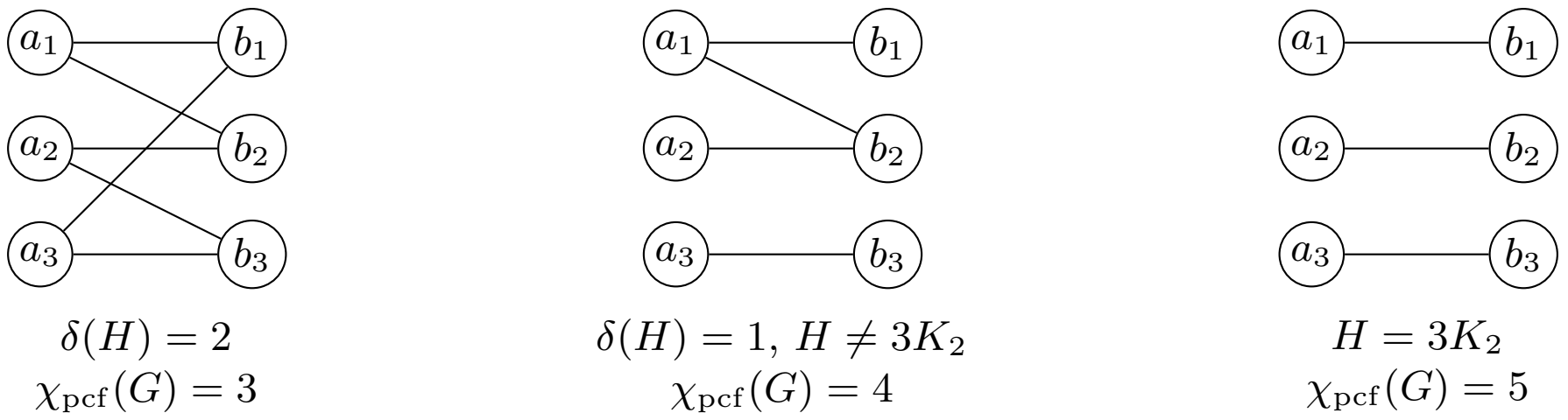

\subsection{Unequal part sizes}

When $a>b$, two additional colors are never needed.  A maximum matching
that fails to saturate $B$ is already handled by
Lemma~\ref{lem:cobip-unsaturated}.  We therefore consider a matching
saturating $B$ and keep $q=a-b\geq1$.

\begin{theorem}
\label{thm:cobip-unbalanced-upper}
If $G$ has a partition into cliques of unequal sizes, then
$\chi_{\mathrm{pcf}}(G)\leq\chi(G)+1$.
A coloring meeting this bound can be constructed in polynomial time.
\end{theorem}

\begin{proof}
Assume $a>b\geq1$.  If $\nu(H)<b$, the earlier lemma gives the stronger
bound $\chi_{\mathrm{pcf}}(G)=\chi(G)$.  Otherwise, let $M$ saturate $B$.
The matching coloring uses $a=\chi(G)$ colors, and its unmatched set
$U\subseteq A$ has size $q\geq1$.

Suppose first that some $y_0\in B$ already satisfies the PCF requirement,
where an isolated vertex counts as satisfying it.  Delete its matching
edge $x_0y_0$.  There are now at least two unmatched vertices in $A$ and
one in $B$.  Every vertex of $A$ sees a private color in $A$, and every
vertex of $B\setminus\{y_0\}$ sees the private color of $y_0$.
The vertex $y_0$ remains valid: before the deletion, its own paired color
was absent from its neighborhood, and splitting this pair into two private
colors changes none of its other neighbor-color counts.
This gives a PCF coloring with $a+1$ colors.

Suppose instead that every vertex of $B$ fails under the original matching
coloring.  By Lemma~\ref{lem:cobip-defects}, each such vertex $y$ has
$N_H(y)=U\cup\{x_y\}$, where $x_yy\in M$.
There must be at least two vertices in $B$.  Indeed, if $B=\{y\}$,
this equality would give $N_H(y)=A$, making $y$ isolated in $G$ and hence
exempt from the PCF condition.

Choose distinct $y_0,y_1\in B$, with partners $x_0,x_1$, and choose
$u\in U$.  Replace the two matching edges $x_0y_0,x_1y_1$ by $uy_1$.
The replacement is allowed because $U\subseteq N_H(y_1)$.
The new matching has size $b-1$, unmatched set
$(U\setminus\{u\})\cup\{x_0,x_1\}$ in $A$, and unmatched vertex $y_0$
in $B$.  As before, the private colors witness all of $A$ and all of
$B\setminus\{y_0\}$.  Also,
$x_1\notin U\cup\{x_0\}=N_H(y_0)$, so $x_1y_0$ is an edge of $G$.
The private color of $x_1$ witnesses $y_0$.  This is again a PCF coloring
with $a+1$ colors.  If one part is empty, $G$ is a clique and needs no additional color.
\end{proof}

To decide whether $a$ colors suffice in the saturated case, we only need
to find a suitable unmatched set.  Call $U\subseteq A$ \emph{feasible}
if $|U|=q$ and $H[A\setminus U,B]$ has a perfect matching.
Define a family $\mathcal F_B$ by taking, for each nonisolated
$y\in B$ with $d_H(y)=q+1$, all sets
$N_H(y)\setminus\{x\}$ with $x\in N_H(y)$.
If $q=1$, also put
$\mathcal F_A=\{\{x\}:x\in A,\ d_H(x)=0\}$; otherwise put
$\mathcal F_A=\varnothing$.
Let $\mathcal F=\mathcal F_A\cup\mathcal F_B$.
Every member has size $q$, and there are at most $|E(H)|+a$ members.

\begin{theorem}
\label{thm:cobip-unbalanced-exact}
Suppose $a>b$ and $\nu(H)=b$.  Then
$\chi_{\mathrm{pcf}}(G)=a$ exactly when there is a feasible unmatched
set $U\notin\mathcal F$.  If there is no such set, then
$\chi_{\mathrm{pcf}}(G)=a+1$.
Whether such a set exists, and a corresponding coloring, can be determined
in polynomial time.
\end{theorem}

\begin{proof}
Every proper $a$-coloring must have exactly $b$ classes of size two, so it
comes from a matching saturating $B$.  By
Lemma~\ref{lem:cobip-defects}, its unmatched set gives a PCF coloring
exactly when it avoids $\mathcal F_B$ and, when $q=1$, avoids
$\mathcal F_A$.  This proves the characterization.  If no such set
exists, the lower bound is $a+1$ and
Theorem~\ref{thm:cobip-unbalanced-upper} supplies that many colors.

It remains to find a feasible set outside the forbidden family without
examining all $q$-subsets of $A$.  We enumerate feasible sets by a binary
search tree over a fixed order of $A$, pruning any partial choice that has
no feasible completion.  A partial choice specifies disjoint sets $U_0,Z_0$
of vertices forced to be unmatched and matched, respectively.  Its
feasibility can be tested by a bipartite perfect-matching computation.

Delete $U_0$ from $A$ and put $A'=A\setminus U_0$.
If $|A'|<b$, reject the partial choice.  Otherwise add
$|A'|-b$ dummy vertices to the $B$ part, each adjacent to every vertex
of $A'\setminus Z_0$.  Keep the original edges of $H[A',B]$.
The two bipartition classes of this auxiliary graph have equal size.
A perfect matching in it matches every original vertex of $B$, and a
vertex of $Z_0$ cannot be matched to a dummy.  Deleting the dummy edges
therefore gives precisely a matching saturating $B$ that also matches
$Z_0$ and leaves $U_0$ unmatched.
Conversely, any such matching can be extended to a perfect matching by
matching its remaining unmatched vertices of $A'$ to the dummies.
This proves that the test is exact.

At each vertex of the fixed order, branch on whether it is in the unmatched
set and keep a branch only if this test succeeds. Each retained node has
a feasible leaf below it. The search has depth $a$. Between consecutive
outputs, a depth-first traversal backtracks along one path and follows the
next, testing at most one rejected sibling at each visited level. Thus it
uses $O(a)$ perfect-matching tests between successive feasible sets. The
same bound applies before the first output and after the last.

Let $F=|\mathcal F|$.  Inspect feasible sets until one lies outside
$\mathcal F$, or until the enumeration ends.  At most $F+1$ feasible
sets need to be inspected, because distinct leaves give distinct sets and
only $F$ sets are forbidden.  Since $F\leq|E(H)|+a$, the number of
matching tests is polynomial.  At a successful leaf, retain its perfect
matching in $H[A\setminus U,B]$ and construct the matching coloring.
\end{proof}

\begin{algorithm}[t]
\DontPrintSemicolon
\KwIn{A co-bipartite graph $G$.}
\KwOut{$\chi_{\mathrm{pcf}}(G)$ and an optimal coloring.}
\If{$V(G)=\varnothing$}{\KwRet{$0$ and the empty coloring}}
find a partition into cliques $A,B$ and arrange $a=|A|\geq b=|B|$\;
\If{$b=0$}{color the clique injectively and \KwRet{$a$ and the coloring}}
\If{$a=b=1$}{solve the two-vertex graph directly and \KwRet{the answer}}
construct the cross-nonedge graph $H$ and a maximum matching $M$\;
\If{$|M|<b$}{
  \KwRet{$a+b-|M|$ and the matching coloring of $M$}\;
}
\If{$a=b$}{
  use Theorem~\ref{thm:cobip-balanced} to return the value and coloring\;
}
put $q\gets a-b$ and construct the forbidden family $\mathcal F$\;
enumerate feasible unmatched sets by the pruned search in
Theorem~\ref{thm:cobip-unbalanced-exact}\;
\ForEach{enumerated feasible set $U$}{
  \If{$U\notin\mathcal F$}{
    find a perfect matching of $H[A\setminus U,B]$\;
    \KwRet{$a$ and its matching coloring}\;
  }
  stop after $|\mathcal F|+1$ feasible sets, or when the search is exhausted\;
}
construct an $(a+1)$-coloring by
Theorem~\ref{thm:cobip-unbalanced-upper}\;
\KwRet{$a+1$ and that coloring}\;
\caption{\textsc{Exact-PCF-Co-Bipartite}$(G)$}
\label{alg:exact-cobipartite}
\end{algorithm}

\begin{corollary}
\label{cor:cobip-algorithm}
The PCF chromatic number and an optimal coloring of an $n$-vertex
co-bipartite graph can be computed in $O(n^6)$ time.
Moreover, $\chi_{\mathrm{pcf}}(G)\leq\chi(G)+1$ unless
$G=\mathrm{CP}_r$ for some $r\geq2$; precisely these exceptional graphs
have $\chi_{\mathrm{pcf}}(G)=\chi(G)+2$.
\end{corollary}

\begin{proof}
The characterization covers equal and unequal part sizes, and
maximum matchings leaving both parts unsaturated. A bipartite matching
can be found by successive augmenting paths in $O(n^3)$ time. Each
feasibility test uses an auxiliary graph with at most $n$ vertices on each
part and at most $n^2$ edges, so this bound applies to every test.
The enumeration uses $O(n)$ tests per feasible set and inspects at most
$|\mathcal F|+1=O(n^2)$ sets. Its total running time is therefore
$O(n^6)$.

The forbidden family contains $O(n^2)$ sets, each with at most $n$
vertices, so it can be written down in $O(n^3)$ time. Comparing one
candidate with every forbidden set takes $O(n^3)$ time, and these checks
across all inspected candidates take $O(n^5)$ time. Finding the final
matching and assigning colors also fit within the $O(n^6)$ bound.
The value statement follows from
Theorems~\ref{thm:cobip-balanced} and~\ref{thm:cobip-unbalanced-upper}.
The path on three vertices shows that the additional color is sometimes
necessary outside the cocktail-party family.
\end{proof}

\begin{corollary}
\label{cor:two-three-cliques}
With a supplied clique partition, PCF Colorability is polynomial-time
solvable for at most two cliques but NP-complete for three cliques.
The hardness for three cliques holds on connected AT-free graphs of
diameter two.
\end{corollary}

\begin{proof}
The two-clique case is Corollary~\ref{cor:cobip-algorithm}.
The graphs in Theorem~\ref{thm:cocomparability-hardness} have a partition into
three cliques and independence number three, so they cannot be covered
by two cliques. They give the second statement.
\end{proof}

\subsection{All graphs with independence number at most two}
\label{subsec:alpha-two}

The matching approach extends beyond co-bipartite graphs.  We now assume
only that $\alpha(G)\leq2$.  This class is AT-free, since it has no
independent triple.  Its complement $H=\overline G$ is triangle-free,
although it need not be bipartite.  Every proper color class still has size
at most two, so proper colorings again correspond to matchings in $H$.
The triangle-free condition makes their PCF constraints simple enough to
check through the unmatched set.

A related matching algorithm of Caro et al. computes the strong odd
chromatic number on this class~\cite[Proposition~29]{CaroEtAlOdd2026}.
Strong odd coloring requires every color present in a neighborhood to
have odd multiplicity. This condition is stricter: the diamond $K_4-e$
needs three PCF colors but four strong odd colors. Our constraints allow
a paired color at a common neighbor, provided another color supplies a
witness. Since every color class has size at most two, PCF coloring on
this class is equivalent to proper odd coloring, which asks for at least
one color of odd multiplicity in each nonempty neighborhood.

We first record a useful observation about the number of colors.  Splitting
a color class into two classes preserves a PCF coloring.  Indeed, if a
vertex had a unique neighbor of the old color, that neighbor retains a
unique color after the split; witnesses of all other colors are unchanged.
Thus a PCF coloring using at most $k$ colors can be refined to use exactly
$k$ colors whenever $k\leq n$.  This observation allows us to fix the
number of edges in the corresponding matching.

Fix an integer $k$ with $\lceil n/2\rceil\leq k\leq n$, and put
$s=n-k$ and $q=2k-n$.  A proper coloring with exactly $k$ nonempty color
classes corresponds to a matching of size $s$ in $H$, with exactly $q$
unmatched vertices.  As before, each matched edge receives one color and
each unmatched vertex receives a private color.

\begin{lemma}
	\label{lem:alpha-two-defects}
	Let $M$ be a matching in the triangle-free graph $H=\overline G$, and
	let $U$ be its unmatched set.  Consider the corresponding matching coloring
	of $G$.  For every nonisolated vertex $v$ of $G$:
	\begin{enumerate}[label=(\roman*),leftmargin=*]
		\item if $v\in U$, then $v$ fails the PCF condition exactly when
		$N_H(v)=U\setminus\{v\}$;
		\item if $v\notin U$, with partner $w$ in $M$, then $v$ fails the PCF
		condition exactly when $N_H(v)=U\cup\{w\}$.
	\end{enumerate}
	In particular, with $q=|U|$, condition (ii) is equivalent to
	$d_H(v)=q+1$ and $U\subseteq N_H(v)$.  Whether the coloring is PCF
	therefore depends only on the unmatched set $U$.
\end{lemma}

\begin{proof}
	Consider a matched edge $xy$ not containing $v$.  Since $H$ is
	triangle-free, $v$ is adjacent in $H$ to at most one of $x,y$.
	Consequently, the pair's color appears once in $N_G(v)$ precisely when
	$v$ is adjacent in $H$ to one endpoint; otherwise it appears twice.
	Also, the private color of an unmatched vertex $u\neq v$ witnesses $v$
	exactly when $u\notin N_H(v)$.
	
	Suppose $v$ is unmatched.  It has no private-color witness exactly when
	$U\setminus\{v\}\subseteq N_H(v)$.  It has no witness from a matched
	pair exactly when it has no $H$-neighbor outside $U$.
	Both statements hold exactly when
	$N_H(v)=U\setminus\{v\}$, proving (i).
	
	Now suppose $vw\in M$.  Its own paired color is absent from $N_G(v)$.
	There is no private-color witness exactly when $U\subseteq N_H(v)$,
	and no witness from another matched pair exactly when all its other
	$H$-neighbors belong to $U$.  These conditions say
	$N_H(v)=U\cup\{w\}$ and prove (ii).
	Finally, if $d_H(v)=q+1$ and $U\subseteq N_H(v)$, there is a unique
	$H$-neighbor outside $U$.  It must be the matching partner of $v$.
	This proves the last assertion.  Isolated vertices of $G$ are exempt
	from all these tests.
\end{proof}

For the fixed value of $q$, form a family $\mathcal F_q$ of forbidden
$q$-subsets of $V(H)$.  For each nonisolated vertex $v$ of $G$ with
$d_H(v)=q-1$, include $N_H[v]$.  For each nonisolated vertex $v$ with
$d_H(v)=q+1$, include every set $N_H(v)\setminus\{w\}$ with
$w\in N_H(v)$.  The first rule contributes at most $n$ sets, and the
second at most $2|E(H)|$.  Duplicate sets may be removed.
Call a $q$-set $U$ feasible if $H-U$ has a perfect matching.

\begin{theorem}
	\label{thm:alpha-two-characterization}
	Let $G$ be an $n$-vertex graph with $\alpha(G)\leq2$, and let
	$\lceil n/2\rceil\leq k\leq n$.  Put $q=2k-n$.
	Then $\chi_{\mathrm{pcf}}(G)\leq k$ if and only if there is a feasible
	$q$-set $U\notin\mathcal F_q$.
\end{theorem}

\begin{proof}
	By refinement, it suffices to consider colorings with exactly $k$ colors.
	Their classes of size two form a matching with $s=n-k$ edges, and the
	unmatched vertices form a feasible $q$-set $U$.  Conversely, a perfect
	matching in $H-U$ gives such a proper coloring.
	
	By Lemma~\ref{lem:alpha-two-defects}, an unmatched vertex $v$ fails
	exactly when $U=N_H[v]$.  Such a set has size $q$ precisely when
	$d_H(v)=q-1$, so it is included by the first rule.
	A matched vertex $v$ fails exactly when $d_H(v)=q+1$ and
	$U=N_H(v)\setminus\{w\}$ for its partner $w$.  These are the sets
	included by the second rule.  Conversely, if a feasible set is included by
	that rule, then $v\notin U$ and its partner must be the unique neighbor
	in $N_H(v)\setminus U$, so the failure really occurs in every matching
	of $H-U$.  We have therefore excluded exactly the unmatched sets whose
	matching colorings fail the PCF condition.
\end{proof}

\begin{lemma}
	\label{lem:alpha-two-enumeration}
	For a fixed matching size $s$, the feasible unmatched sets in a graph $H$
	can be enumerated with polynomial delay. More precisely, $O(n)$ general
	perfect-matching tests suffice between consecutive outputs.
\end{lemma}

\begin{proof}
	Use a fixed order of $V(H)$ and branch on whether each vertex is unmatched.
	A partial choice specifies disjoint sets $U_0,Z_0$ forced to be unmatched
	and matched.  We show how to test whether a matching of size $s$ can satisfy
	these requirements.
	
	Delete $U_0$ and let $n'=n-|U_0|$.  If $n'<2s$, the partial choice is
	impossible.  Otherwise add $d=n'-2s$ new dummy vertices.  Make them
	independent and join each of them to every vertex of
	$V(H)\setminus(U_0\cup Z_0)$; retain all edges of $H-U_0$.
	The resulting graph has $n'+d=2(n'-s)$ vertices.
	
	A perfect matching in this graph matches all $d$ dummies to distinct
	original vertices outside $Z_0$.  Removing these dummy edges leaves
	exactly $s$ edges of $H$, covering $Z_0$ and avoiding $U_0$.
	Conversely, any size-$s$ matching with those properties leaves exactly
	$d$ vertices of $H-U_0$ unmatched, all outside $Z_0$.  Matching them to
	the dummies completes a perfect matching of the auxiliary graph.
	Thus one general perfect-matching test decides whether the partial choice
	has a completion.
	
	Keep a search-tree node only when this test succeeds. Each retained node
	has a feasible leaf below it. The search has depth $n$. Between consecutive
	outputs, a depth-first traversal backtracks along one path and follows the
	next, testing at most one rejected sibling at each visited level. It
	therefore makes $O(n)$ tests between outputs, and also before the first
	output and after the last. At a leaf, the unmatched set has size $n-2s$,
	and its remaining vertices have a perfect matching. The fixed vertex order
	ensures that no unmatched set is output twice.
\end{proof}

\begin{theorem}
	\label{thm:alpha-two-algorithm}
	For graphs with independence number at most two, PCF Colorability can be
	decided in $O(n^6)$ time. The PCF chromatic number and an optimal coloring
	can be computed in $O(n^6\log n)$ time.
\end{theorem}

\begin{proof}
	Handle the empty graph directly.  For a nonempty graph, $k<n/2$ is
	impossible because a proper color class has size at most two, while
	$k\geq n$ always suffices.  In the remaining range, construct
	$H=\overline G$, put $s=n-k$, $q=2k-n$, and form $\mathcal F_q$.
	
	Apply Lemma~\ref{lem:alpha-two-enumeration} and inspect feasible sets
	until one lies outside $\mathcal F_q$, or the enumeration ends.
	At most $|\mathcal F_q|+1$ distinct feasible sets need to be inspected.
	If that many exist, at least one is outside the forbidden family.
	Otherwise the enumeration ends after listing all feasible sets.
	Since $|\mathcal F_q|\leq n+2|E(H)|=O(n^2)$, the enumeration makes
	$O(n^3)$ perfect-matching tests. Each auxiliary graph has at most $2n$
	vertices and $O(n^2)$ edges. Polynomial-time general matching goes back to
	Edmonds~\cite{Edmonds1965}. For the running-time bound, we may use the
	$O(\sqrt{|V|}|E|)$ algorithm of Micali and
	Vazirani~\cite{MicaliVazirani1980}, which takes $O(n^{5/2})$ time, and
	hence at most $O(n^3)$ time, on these auxiliary graphs. The total time for
	the matching tests is therefore $O(n^6)$.
	
	Writing down the forbidden family takes $O(n^3)$ time. Comparing one
	candidate with every forbidden set takes $O(n^3)$ time, so all membership
	checks take $O(n^5)$ time. These steps fit within the decision bound.
	
	At an accepting leaf, retain a perfect matching of $H-U$ and give its
	pairs and the unmatched vertices distinct colors.  The characterization
	proves that this is a PCF coloring. Finally, feasibility is monotone in
	$k$, so binary search over $\lceil n/2\rceil,\ldots,n$ uses
	$O(\log n)$ decision calls. Recovering a coloring at the smallest accepted
	value gives an optimal coloring within $O(n^6\log n)$ time.
\end{proof}

\begin{algorithm}[t]
	\DontPrintSemicolon
	\KwIn{A graph $G$ with $\alpha(G)\leq2$, and an integer $k$.}
	\KwOut{A PCF coloring using at most $k$ colors, or \textnormal{\textsc{No}}.}
	\If{$V(G)=\varnothing$}{\KwRet{the empty coloring if $k\geq0$, otherwise \textnormal{\textsc{No}}}}
	\If{$k<\lceil |V(G)|/2\rceil$}{\KwRet{\textnormal{\textsc{No}}}}
	\If{$k\geq |V(G)|$}{give every vertex a different color and \KwRet{the coloring}}
	put $n\gets |V(G)|$, $H\gets\overline G$, $s\gets n-k$, $q\gets2k-n$\;
	construct the forbidden family $\mathcal F_q$\;
	enumerate feasible unmatched $q$-sets using
	Lemma~\ref{lem:alpha-two-enumeration}\;
	\ForEach{enumerated set $U$}{
		\If{$U\notin\mathcal F_q$}{
			find a perfect matching of $H-U$\;
			give each matching edge one color and each vertex of $U$ a private color\;
			\KwRet{the resulting coloring}\;
		}
		stop after $|\mathcal F_q|+1$ sets, or when the search is exhausted\;
	}
	\KwRet{\textnormal{\textsc{No}}}\;
	\caption{\textsc{PCF-Alpha-Two}$(G,k)$}
	\label{alg:alpha-two}
\end{algorithm}

\begin{corollary}
	\label{cor:alpha-two-three-boundary}
	PCF Colorability is polynomial-time solvable when $\alpha(G)\leq2$,
	but NP-complete when $\alpha(G)\leq3$.  The hardness at independence
	number three holds even for connected cocomparability graphs of diameter
	two, with a supplied partition into three cliques and target
	$k=\chi(G)=\omega(G)$.
\end{corollary}

\begin{proof}
	The positive statement is Theorem~\ref{thm:alpha-two-algorithm}.
	The graphs in Theorem~\ref{thm:cocomparability-hardness} have independence number
	exactly three and satisfy all the additional restrictions.
\end{proof}

\section{Exact coloring on distance-hereditary graphs}
\label{sec:dh}

A graph is distance-hereditary if every connected induced subgraph preserves
the distances between its vertices. Distance-hereditary graphs include trees
and complete bipartite graphs, but their PCF chromatic number is unbounded
even on the bipartite subclass. This follows, for example, from the open-neighborhood
conflict-free lower bound of Bhyravarapu et
al.~\cite{BhyravarapuEtAl2024}.
Consequently, testing only a fixed number of colors cannot give an exact
algorithm for the whole class.

We give an algorithm that permits the number of colors to grow with the
input. At each decomposition node, the algorithm records how many colors
have each of five possible behaviors. It also records the smallest palette
that realizes these counts. The resulting algorithm computes an optimal
coloring in $O(n^{11}+m)$ time, without assuming a bound on the clique
number or on the answer.

\subsection{The decomposition and its exposed vertices}
\label{subsec:dh-decomposition}

We use a binary decomposition whose leaves are the vertices of $G$. At a
node $t$, let $G_t$ be the graph on the leaves below $t$, and let
$T_t\subseteq V(G_t)$ be its designated twin set. A leaf $v$ represents
$G_t=K_1$ with $T_t=\{v\}$. For an internal node with children $t_1,t_2$,
write $T_1=T_{t_1}$ and $T_2=T_{t_2}$. There are three operations:

\begin{itemize}
\item A \emph{true-twin operation} adds every edge between $T_1$ and $T_2$
and sets $T_t=T_1\cup T_2$.
\item A \emph{false-twin operation} adds no edge between the children and
sets $T_t=T_1\cup T_2$.
\item An \emph{attachment operation} adds every edge between $T_1$ and
$T_2$ and sets $T_t=T_1$. This operation is ordered: the first child is
retained, while the second child becomes internal at the parent.
\end{itemize}

These operations give the standard decomposition of distance-hereditary
graphs; a decomposition with $O(n)$ nodes can be constructed in $O(n+m)$
time~\cite{HsiehHoHsuKo2006}. They are also consistent with the classical
construction by pendant vertices and true or false twins
\cite{BandeltMulder1986}. To see the connection, start with one vertex leaf.
When a new vertex is added as a twin of $v$, replace the leaf $v$ by the
corresponding twin operation on the old and new leaves. For a pendant
addition, use an attachment node that retains the old leaf. Thus the new
pendant vertex inherits no edge outside this replacement. The other two
operations make the new vertex inherit exactly the old vertex's neighbors,
with an edge between the twins precisely in the true-twin case.
Disconnected components may be combined by false-twin operations.

The name twin set does not mean that its vertices have the same neighbors
inside $G_t$. What matters is their common neighborhood outside $G_t$.

\begin{lemma}
\label{lem:dh-interface}
For each node $t$, every vertex of $T_t$ has the same neighborhood outside
$G_t$, and no vertex of $V(G_t)\setminus T_t$ has a neighbor outside $G_t$.
\end{lemma}

\begin{proof}
At the root there are no outside vertices. Now consider the children of a
node for which the assertion holds. Edges between the children join either
all of their designated twin sets or none of them. A child twin set is
either included in the parent's twin set or has no neighbors outside the
parent graph. In each case, all vertices of the child twin set acquire the
same outside neighbors. A vertex outside that twin set acquires neither an
edge to the other child nor an edge outside the parent. Applying this
argument from the root down proves the assertion.
\end{proof}

We will call $T_t$ the \emph{exposed set}. It contains all vertices whose
witnesses may still depend on colors outside the current subgraph.

\subsection{Partial colorings with prescribed outside counts}
\label{subsec:dh-promises}

Write $\tau(a)=\min\{2,a\}$. Fix a palette $[k]$, which may contain unused
colors. For each color $i$, record two values $m_i,q_i\in\{0,1,2\}$.
The value $m_i$ is the number of vertices of color $i$ in $T_t$, capped
at two. The value $q_i$ prescribes the number of neighbors of color $i$
outside $G_t$ seen by each exposed vertex, again capped at two.
The vector $q$ is a promise about outside contributions; it need not be
realizable by an outside graph when the state is considered on its own.
The merge rules will make these promises consistent.

We define feasibility precisely. A proper coloring $c:V(G_t)\to[k]$
realizes $(m,q)$ if its exposed counts are $m$, if $q_{c(v)}=0$ for every
$v\in T_t$, and if it satisfies the following witness test. For a vertex
$v$, let $a_v(i)$ be the capped number of its neighbors of color $i$ in
$G_t$, adding $q_i$ when $v\in T_t$. If all values $a_v(i)$ are zero,
then $v$ is treated as isolated and needs no witness. Otherwise, some
$a_v(i)$ must equal one. Vertices outside $T_t$ receive no outside
contribution. At the root, where $q=0$, this is exactly the PCF condition,
including the exemption for isolated vertices.

The condition $q_{c(v)}=0$ enforces properness against the promised outside
neighbors. It implies that $m_i$ and $q_i$ cannot both be positive.
There are therefore only five possible pairs for one color:
$\mathcal P=\{(0,0),(1,0),(2,0),(0,1),(0,2)\}$.
Their meanings are summarized below. A value two always means ``at least
two,'' because all counts are truncated.

\begin{center}
\begin{tabular}{c|l}
pair & meaning at the exposed set\\
\hline
$(0,0)$ & absent from the exposed set and from its outside neighbors\\
$(1,0)$ & occurs once in the exposed set\\
$(2,0)$ & occurs at least twice in the exposed set\\
$(0,1)$ & seen once outside by every exposed vertex\\
$(0,2)$ & seen at least twice outside by every exposed vertex
\end{tabular}
\end{center}
A feasible state must test all vertices of $G_t$ under its promise. Merely
recording the colors on $T_t$ would not preserve enough information about
witnesses.

The phrase ``isolated under the promise'' is local to a state. In every
state used by a reconstructed root solution, the ancestors realize the
promised counts exactly, so the local test agrees with isolation in the
completed graph. At the root, the promise is zero, and all counts are
actual counts.

At a leaf, exactly one color has $m_i=1$ and all others have $m_i=0$.
The chosen color must have $q_i=0$. The witness test succeeds if some
outside count is one, or if all outside counts are zero and the vertex is
isolated under the promise.

\subsection{How the promises combine}
\label{subsec:dh-labeled-merge}

For an internal node, let $e$ indicate whether edges are added between the
child exposed sets. Let $\rho_j$ indicate whether $T_j$ is retained in
$T_t$. The three operations have the following parameters:

\begin{center}
\begin{tabular}{l|ccc}
operation & $e$ & $\rho_1$ & $\rho_2$\\
\hline
true twin & $1$ & $1$ & $1$\\
false twin & $0$ & $1$ & $1$\\
attachment & $1$ & $1$ & $0$
\end{tabular}
\end{center}

Consider one color. Suppose its exposed counts in the two children are
$a,b$, and its promised count outside the parent is $q$. Its parent count
$m$ and child promises $q_1,q_2$ are then
\begin{equation}
\label{eq:dh-merge}
 m=\tau(\rho_1a+\rho_2b),\qquad
 q_1=\tau(\rho_1q+eb),\qquad
 q_2=\tau(\rho_2q+ea).
\end{equation}
For example, an exposed vertex of the first child sees the second child's
$b$ vertices of this color when $e=1$. It sees the parent's outside
neighbors only when $\rho_1=1$. These are the two terms in $q_1$.
Capping before or after addition gives the same result, so counts above
two do not have to be distinguished.

Call a triple $(a,b,q)\in\{0,1,2\}^3$ \emph{allowed} when all three
pairs $(a,q_1),(b,q_2),(m,q)$ produced by
\eqref{eq:dh-merge} belong to $\mathcal P$. Let $\mathcal A_t$ be the
set of allowed triples for the operation at $t$.

\begin{lemma}
\label{lem:dh-labeled-merge}
A parent profile $(m,q)$ is feasible on a fixed labeled palette if and only
if child exposed counts can be chosen so that the two child profiles are
feasible and every color satisfies \eqref{eq:dh-merge}.
\end{lemma}

\begin{proof}
First restrict a feasible parent coloring to its children. By
Lemma~\ref{lem:dh-interface}, the neighbors outside either child consist
of the indicated opposite exposed set, together with the parent's outside
neighbors when that child's exposed set is retained. Thus their color
counts are exactly $q_1,q_2$. Every vertex is tested against the same
complete neighborhood counts as in the parent state, so both child
profiles are feasible.

Conversely, take feasible child colorings with the indicated labeled
profiles. They are proper within the children. If $e=1$, a color cannot
appear in both exposed sets: $a>0$ and $b>0$ would give $q_1>0$ and
violate properness of the first child. Hence all added edges are proper.
The condition $(m,q)\in\mathcal P$ also enforces properness against the
parent promise.

For an exposed vertex in either child, its child promise is exactly the
contribution from the other child and from outside the parent. For every
other vertex, neither contribution is present. The child witness tests
therefore remain valid in the parent. This also preserves the isolated
case: a zero neighborhood count stays zero precisely when no neighbor has
been added. The combined coloring realizes the parent profile.
\end{proof}

\subsection{Recording counts of color types}
\label{subsec:dh-histograms}

The names of the colors are irrelevant. For each pair $p\in\mathcal P$,
let $h_p$ be the number of colors whose pair equals $p$. We call these five
numbers a \emph{histogram}. They sum to $k$, so there are
$\binom{k+4}{4}=O(k^4)$ possible histograms.

\begin{lemma}
\label{lem:dh-permutation}
On a palette of size $k$, feasibility depends only on the histogram of the
pairs $(m_i,q_i)$.
\end{lemma}

\begin{proof}
If two profiles have the same histogram, match their color labels within
each pair type. This gives a permutation of the palette that maps one
profile to the other. Apply the same permutation to every vertex color
and to the promise. Properness and the condition that a neighbor color
occurs once are unchanged. Thus either both profiles are feasible or
neither is feasible.
\end{proof}

A merge can now be specified by a nonnegative integer $z_{a,b,q}$ for
each allowed triple. This number tells us how many colors have that
triple. The triple counts determine the parent histogram and both child
histograms through \eqref{eq:dh-merge}. If the two child histograms are
feasible, they can be realized on the same palette. Indeed, first assign
the labeled colors to the indicated triples. Independently permute a
feasible coloring of each child so that its pairs match this assignment.
Lemma~\ref{lem:dh-permutation} permits both permutations. The preceding
merge lemma then combines the children. Conversely, every feasible parent
coloring determines these triple counts.

This argument keeps the witness requirements of all exposed vertices.
It does not assume that two colorings with the same boundary counts have
the same witnesses. Rather, feasibility means that some coloring satisfies
all vertex requirements under the entire promised vector $q$.

The number of allowed triples is small. At a false-twin node, either
$q=0$ and $a,b$ are arbitrary, or $q>0$ and $a=b=0$. This gives eleven
triples. At a true-twin node, $q=0$ permits at most one of $a,b$ to be
positive, and $q>0$ again forces both to be zero. This gives seven.
At an attachment node, either $a>0$ and $b=q=0$, or $a=0$ and $b,q$ are
arbitrary. This gives eleven. The triple $(0,0,0)$ is allowed in all cases.

\subsection{Computing the minimum palette in one pass}
\label{subsec:dh-minimum-palette}

The fixed-palette histograms already give a polynomial decision algorithm.
We can also optimize directly, without repeating the algorithm in a binary
search. The key is to omit the count $h_{00}$ and keep the smallest palette
that supplies the omitted colors.

Fix a budget $K\geq1$. For a four-tuple
$h=(h_{10},h_{20},h_{01},h_{02})$ with nonnegative entries and sum at most
$K$, let $D_t(h)$ be the smallest palette size $k\leq K$ for which a
feasible profile has these four counts. Set $D_t(h)=\infty$ if no such
profile exists. This is the size of the whole palette, including colors
that occur only in the interior of $G_t$ or are required by its outside
promise. It is not just the number of colors on $T_t$.

Adding a new unused color with pair $(0,0)$ preserves feasibility.
Consequently, if $D_t(h)=k$, the same four-tuple can be realized on every
palette of size between $k$ and $K$. This padding property is what allows
us to optimize the palette while keeping only four counts.

At a leaf, the finite entries are
$D_t(1,0,b,c)=1+b+c$ for $1+b+c\leq K$, provided either $b\geq1$ or
$b=c=0$. The exposed vertex needs one color. The $b+c$ colors with positive
outside counts are different from that color. If $b\geq1$, an outside
color occurs once and supplies a witness. If $b=c=0$, the vertex is
isolated. The remaining case, $b=0<c$, has neighbors but no unique color
and is rejected.

For an internal node, put
$\mathcal A_t^+=\mathcal A_t\setminus\{(0,0,0)\}$.
Enumerate nonnegative counts $z$ on $\mathcal A_t^+$ with total
$L(z)\leq K$. Their projections determine four-tuples $h_1(z),h_2(z)$
for the children and $h(z)$ for the parent; a projected pair $(0,0)$ does
not contribute to these four-tuples. The recurrence is
\begin{equation}
\label{eq:dh-minimum-recurrence}
 D_t(h)=\min_{z:\,h(z)=h}
 \max\bigl\{D_{t_1}(h_1(z)),D_{t_2}(h_2(z)),L(z)\bigr\},
\end{equation}
where we ignore infinite candidates and candidates exceeding $K$.

\begin{lemma}
\label{lem:dh-minimum-recurrence}
The initialization and recurrence compute the stated values of $D_t$.
\end{lemma}

\begin{proof}
The leaf initialization follows from its witness test. Assume that both
child tables are correct, and consider one finite candidate in
\eqref{eq:dh-minimum-recurrence}. Let its value be $k$.
Pad a feasible coloring for each child to a palette of size $k$.
The $L(z)$ nonempty triples prescribe that many labeled colors. Give each
of the other $k-L(z)$ colors the empty triple $(0,0,0)$.

The resulting full child histograms have the required four nonzero counts
and exactly the remaining number of $(0,0)$ colors. In a child, some
nonempty parent triples may also project to $(0,0)$. They are included in
that remaining number; they are not lost when the empty triples are added.
By Lemma~\ref{lem:dh-permutation}, we can independently permute the two
child colorings to match these full histograms. The labeled merge lemma
then gives a feasible parent coloring on $k$ colors. Thus every candidate
is achievable.

Conversely, take a feasible parent profile on $k\leq K$ colors. Its labeled
triples are allowed by Lemma~\ref{lem:dh-labeled-merge}. Remove the empty
triples and call the remaining counts $z$. There are at most $k$ nonempty
triples, so $L(z)\leq k$. The restrictions to both children realize their
projected four-tuples on the same palette. Hence their minimum palette
sizes are at most $k$. The recurrence therefore considers a candidate of
value at most $k$ for this parent four-tuple. Together with achievability,
this proves equality.
\end{proof}

At the root $r$ there are no outside neighbors, so we require
$h_{01}=h_{02}=0$. The least finite value
$D_r(h_{10},h_{20},0,0)$ is the minimum palette size up to the budget.
If all such entries are infinite, no PCF coloring within the budget exists.

\begin{algorithm}[t]
\DontPrintSemicolon
\caption{\textsc{Minimum-PCF-DH}$(\mathcal T,K)$}
\label{alg:dh-pcf}
\KwIn{A binary distance-hereditary decomposition $\mathcal T$ and a positive
palette budget $K$.}
\KwOut{A minimum PCF coloring within the budget, or a report that none exists.}
\ForEach{node $t$ in postorder}{
 Initialize $D_t(h)\gets\infty$ for every four-tuple $h$ with sum at most $K$\;
 \eIf{$t$ is a leaf}{
  \ForEach{$b,c\geq0$ with $1+b+c\leq K$}{
   \If{$b\geq1$ or $b=c=0$}{set $D_t(1,0,b,c)\gets1+b+c$\;}
  }
 }{
  Form the constant set $\mathcal A_t^+$ from the operation at $t$\;
  \ForEach{nonnegative integer vector $z$ on $\mathcal A_t^+$ with
  $L(z)\leq K$}{
   Compute its child projections $h_1,h_2$ and parent projection $h$ using
   \eqref{eq:dh-merge}\;
   \If{$D_{t_1}(h_1)$ and $D_{t_2}(h_2)$ are finite}{
    $k\gets\max\{D_{t_1}(h_1),D_{t_2}(h_2),L(z)\}$\;
    \If{$k\leq K$ and $k<D_t(h)$}{
     Set $D_t(h)\gets k$ and store $z$ as its predecessor choice\;
    }
   }
  }
 }
}
Choose a root tuple $h=(h_{10},h_{20},0,0)$ of minimum finite value\;
\If{no such tuple exists}{\KwRet{no PCF coloring uses at most $K$ colors}\;}
Set $k^*\gets D_r(h)$ and reconstruct a coloring on $[k^*]$ from the stored
choices\;
\KwRet{the coloring and $k^*$}\;
\end{algorithm}

\subsection{Running time and coloring reconstruction}
\label{subsec:dh-complexity}

There are $\binom{K+4}{4}=O(K^4)$ possible four-tuples at a node.
A false-twin or attachment node has ten nonempty triple types, so it
considers $\binom{K+10}{10}=O(K^{10})$ transition vectors whose total is at
most $K$. A true-twin node has six nonempty triple types and considers
$\binom{K+6}{6}=O(K^6)$ vectors. For each vector, the projections and
candidate value take constant time because the number of types is fixed.
Arithmetic on these counts and array lookups by the four coordinates take
constant time in the standard word-RAM model. Initializing a table takes
$O(K^4)$ time, and initializing its finite leaf entries takes $O(K^2)$
time. The binary decomposition has $O(n)$ nodes, so the dynamic program
takes $O(nK^{10})$ time. We retain one successful predecessor vector $z$
for each finite entry to recover a coloring.

For completeness, the stored choices recover actual color labels. Start
with a minimum root entry and a palette $[k^*]$. Assign these labels to
its pair types, giving every unused position the type $(0,0)$.
At an internal node, its stored vector $z$ prescribes the numbers of labels
in each nonempty triple. Within each of the five parent pair types,
including the implicit $(0,0)$ type, partition the currently assigned
labels according to these triples. Give the remaining
$k^*-L(z)$ labels the empty triple. This is possible because $z$ projects
to the parent four-tuple. Equation~\eqref{eq:dh-merge} now supplies labeled
profiles for both children. Continue recursively. At a leaf, use its sole
color with exposed count one.

Although a child table entry may have been obtained with fewer than $k^*$
colors, padding permits it to use the same global palette during recovery.
No additional graph choices have to be guessed. The proof of
Lemma~\ref{lem:dh-minimum-recurrence} guarantees that the recovered labels
satisfy every promise and every vertex witness test. Recovery takes
$O(nk^*)$ time and is included in the dynamic-programming bound.

\begin{theorem}
\label{thm:dh-exact}
Let $G$ be a nonempty $n$-vertex distance-hereditary graph with $m$ edges.
Given its binary decomposition, PCF $k$-colorability can be decided and a
coloring recovered in $O(nk^{10})$ time for $1\leq k\leq n$. For any
positive $k$, the same bound holds with $k$ replaced by $\min\{k,n\}$.
Constructing the decomposition from an explicit graph adds $O(n+m)$ time.
An optimal PCF coloring can be constructed in $O(n^{11}+m)$ time from an
explicit graph input.
\end{theorem}

\begin{proof}
For the decision problem, run Algorithm~\ref{alg:dh-pcf} with budget
$K=\min\{k,n\}$. Distinct colors on all vertices give a PCF coloring, so
replacing a larger budget by $n$ does not change the answer. The minimum
root entry is finite exactly when a PCF coloring within the budget exists.
For optimization, use $K=n$ in a single run. This budget contains an
optimum, and the recurrence computes it directly without a binary search.
The stated bounds follow from the transition count, the reconstruction
procedure, and the $O(n+m)$ decomposition construction.
\end{proof}

The algorithm treats an unused color and a color used only in the interior
as the same pair $(0,0)$. This does not assume that those situations have
identical vertex witnesses. Their possible effects on witnesses are
already included in the feasibility of the state under its promised
outside vector. The permutation and padding arguments are what make this
compression valid.

\section{Conclusion and future research}
\label{sec:conclusion}

PCF Colorability is NP-complete on cocomparability graphs, even when an
optimal ordinary coloring and a dominating edge are supplied. Every graph
in the reduction needs either $\omega(G)$ or $\omega(G)+1$ colors.
The result also gives hardness on AT-free graphs. Within the
cocomparability class, co-bipartite graphs admit a complete matching
characterization and a polynomial-time algorithm. The matching method
extends to all graphs with independence number at most two, whereas the
hard instances have independence number three. Thus both the number of
clique parts and the independence number reveal a tractability boundary.

For distance-hereditary graphs, the decomposition allows us to group
colors by their effect on the exposed vertices. Counts truncated at two
preserve exactly the information needed for a witness. The resulting
histogram has five possible color types and supports exact optimization
without fixing the number of colors.

Several questions remain. Our hardness reduction uses a palette that
grows with the input, so the complexity on AT-free graphs for each fixed
number of colors is a separate question. Finding a polynomial-time
algorithm for every fixed palette, or proving hardness for a particular
fixed value, would clarify this point. It would also be useful to identify
further tractable cocomparability subclasses beyond co-bipartite graphs,
especially classes that allow independence number three but restrict how
the clique parts interact.

The distance-hereditary algorithm takes $O(n^{11}+m)$ time. Its
recurrence leaves room for improvement.
Many histograms considered by the algorithm may never be reached. A
structural description of the reachable states could simplify the
transitions. The true-twin, false-twin, and attachment operations may also
admit more direct combinations than the common recurrence used here.
Finally, the method suggests studying other local coloring conditions
whose neighborhood counts can be truncated at a fixed value. Each such
extension would need a proof that grouping colors preserves all the
required information.

\bibliographystyle{plain}
\bibliography{PCF_ref}
         
\end{document}